# ABALONE[TM] Photosensors for the IceCube Experiment

Daniel Ferenc[1,2]
Andrew Chang[1], Cameron Saylor[1]
Sebastian Böser[3]
Alfredo Davide Ferella[4]
Lior Arazi[5]
John R. Smith[2], Marija Šegedin Ferenc[2]

(1) University of California, Davis, CA, USA
(2) PHOTONLAB, INC., Davis, CA, USA
(3) Johannes Gutenberg University, Mainz, Germany
(4) Stockholms Universitet, Stockholm, Sweden
(5) Nuclear Engineering Unit, Faculty of Engineering Sciences, Ben-Gurion University of the Negev, Israel

ABSTRACT

The ABALONE[TM] Photosensor Technology (U.S. Pat. 9,064,678) is a modern, scalable technology specifically invented for cost-effective mass production, robustness, and high performance. We present the performance of advanced fused-silica ABALONE Photosensors, developed specifically for the potential extension of the IceCube neutrino experiment, and stress-tested for 120 days. The resulting performance makes a significant difference: intrinsic gain in the high $10^8$ range, total afterpulsing rate of only $5 \times 10^{-3}$ ions per photoelectron, sub-nanosecond timing resolution, single-photon sensitivity, and unique radio-purity and UV sensitivity, thanks to the fused silica components—at no additional cost to the assembly process.

1. INTRODUCTION

To this day, the broad field of large-area photon detection has relied on the expensive (more than $100,000/m$^2$) 80-year-old Photomultiplier Tube (PMT) technology, while the millimeter-sized all-silicon devices, particularly the Geiger-Mode Avalanche Photo Diodes (G-APDs, SiPMs), still remain too expensive for most of large-area applications (currently around $10 million/m$^2$). Our studies have led to the conclusion that the reason why the PMT and similar technologies could never have evolved into a modern and cost-effective vacuum technology (such as the production of blank compact discs (CDs) (< $10/m$^2$), or plasma TV screens (< $500/m$^2$)), lies precisely in the combination of materials that the PMTs, Hybrid Photo-Diodes (HPDs) and Microchannel Plate PMTs (MCP-PMTs) are made of. The components comprising those materials – intrinsic to their very concepts – include glass, combined with metals, wire feedthroughs, brazes, spot-welds, ceramics, alkali metal deposition sources, heating wires, platinum-antimony beads, microchannel plates, photodiodes, or silicon photomultipliers. This disparate combination of materials inevitably necessitates a daylong, static, 'batch-mode' vacuum processing, which precludes any modern continuous-line production methods. Despite some recent advances in these technologies [1-5], they remain subject to this fundamental limitation

In contrast, the ABALONE Photosensor concept [6,7] has bypassed that critical manufacturing problem through the—highly non-trivial—avoidance of all materials other than glass (or a similar dielectric material), among the components entering the uninterrupted vacuum production line. This 'glass-only' concept draws from the fact that, unlike metals, glass does not release trapped gases from its bulk at the temperatures of interest [8]. Therefore, it does not need a deep bakeout; instead, it requires only a thorough surface-layer 'scrub'. The ABALONE Photosensor design comprises only three industrially pre-fabricated glass components (Fig.1) that enter a continuous production-line process, which starts with standard rapid plasma cleaning, followed by a standard, rapid ultrapure-metal thin film deposition process, and ends with hermetic bonding of the three components, as the two thin-film seals between them are activated [6,9].

The ABALONE Photosensor prototypes have been developed and initially produced in the specially constructed ABALONE Prototype Pilot Production Plant (A4P) (part of the Ferenc-Lab at UC Davis), which is in essence a downscaled rudimentary version of a real production line that nevertheless performs all of the processes accurately [6]. Based on that R&D, PhotonLab, Inc. has recently designed and acquired a production-oriented facility, in order to perform application-specific R&D, and gradually scale-up production. One of the keys to this technology is our thin-film vacuum-sealing technique [6,9]. It was optimized through several generations of full-scale prototypes. Verifying lasting integrity of any vacuum-sealed photosensor is a complex task that can be properly carried out only over a long period, and only in a permanent active detection mode. The continuous, 4-year long test of one of the ABALONE Photosensor prototypes that was left working on the test bench ever since it was assembled in May of 2013, has revealed [10] the afterpulsing rate of positive ions two orders of magnitude lower than in PMTs [5], and constantly improving with time. The low intrinsic production cost of the ABALONE Photosensors (expected to fall below \$5,000/m$^2$ with fully developed production) is also accompanied by simple, application-specific, cost-effective integration methods [6,7]. For example, ABALONE Photosensors can be integrated within a Tandem detector module that hosts two ABALONE units oriented in opposite directions [6], or, quite differently, in thin, lightweight, self-supporting and modular composite ABALONE-Panels [7]. The Tandem module configuration has been specifically developed for the IceCube extension project. The ABALONE Photosensors are resistant to shock, vibration and compression, and immune to accidental exposure to strong light, including daylight.

This article focuses on the first stress tests of the new generation of fused-silica ABALONE Photosensors, specifically developed for the IceCube extension project, conducted at the extreme voltage and event rate settings. One of the prototypes has also been tested for integrity at a temperature of -56℃, representative of the IceCube environment. In parallel, we have also been developing ABALONE Photosensor Panels specifically designed for cryogenic dark matter and double-beta decay experiments. The results of the various specific designs and optimizations will appear in the forthcoming articles.

## 2. ABALONE PHOTOSENSORS

Each photoelectron emerging from the photocathode on the inside surface of the Dome, see Fig.1, is accelerated into the small hole in the center of the Base Plate, which is vacuum-sealed by the Windowlet from the outside. The Windowlet-scintillator converts the kinetic energy of an accelerated photoelectron into secondary photons, proportionally to the kinetic energy and the scintillator yield [11,12]. In the currently presented setup, a 25 keV photoelectron generates around 650 photons in the LYSO scintillator, and 100 of those are detected by the SensL J-type G-APD with an internal gain of $\approx 6.1 \times 10^6$, so the combined gain is $\approx 6 \times 10^8$.

The breakthrough concept of the two sealing thin films doubling as passes for the high voltage and the ground potential [6] allows the ABALONE Photosensor to function without any through-the-glass feedthroughs, i.e. without any solid metals among the vacuum-processed components. The absence of metals and other non-glass materials presents the key to the modern ABALONE technology. The electron-focusing field within the vacuum enclosure can be formed in different suitable ways, depending on the desired integration configuration [6,7], by the shape of the conductor, kept at the same potential as the photocathode, positioned below the base Plate (not shown here). That conductor assures collection of electrons from the entire photocathode, while in its absence the collection efficiency of electrons originating from the bottom 5-10 mm of the Dome falls below unity. The concentration factor of 10,000 from the photocathode to the bombarded Windowlet area, minimizes the G-APD area and thus the overall cost, as well as its capacitance. The Windowlet can either be entirely made of a suitable scintillator material, or as a thin glass plate coated on its vacuum-facing surface with a thin scintillator film.

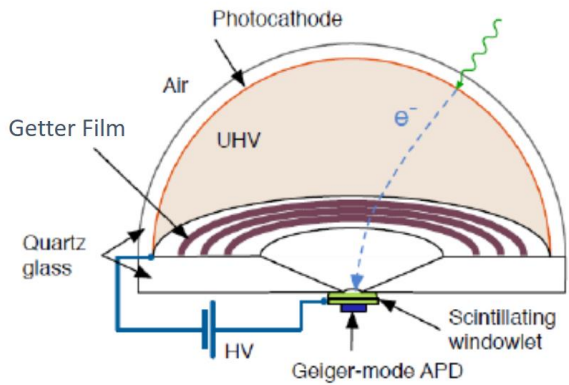
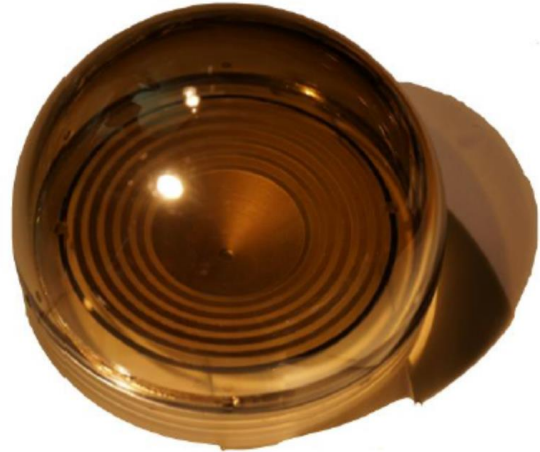

Fig. 1 Left: The ABALONE[TM] Photosensor schematics. Right: A prototype developed for the IceCube extension project. The inner diameter of the Dome is 10 cm. The base plate and the Dome are made of fused silica, which provides UV sensitivity, as well as a unique level of radio purity. The ABALONE Technology allows fused silica to be used instead of glass without any additional cost to the assembly process.

3. EXPERIMENTAL

For the ABALONE prototype discussed in this article, we have used a $Cs_3Sb$ photocathode, while the Windowlet was made of LYSO scintillator. Any other bialkali, multialkali or similar photocathode materials can be grown in the Dome under the controlled conditions within the production system, while the $Cs_3Sb$ photocathode is most suitable for R&D since during the deposition process it spontaneously converges to the optimal stoichiometric configuration [10,13-16]. The Windowlet is a 6x6x1.5 $mm^3$ plate. We chose LYSO for this study because of its mechanical and chemical robustness. Scintillators that offer higher yields and faster response than LYSO have been studied recently, but LYSO already provides an acceptable performance for most applications considered so far. A dismountable optical coupling of the G-APD to the Windowlet was established with standard optical grease. The presented data were collected exclusively with a 6x6 $mm^2$ SensL J-type MicroFJ G-APD that comprises 35μm cells. The combined gain of the LYSO Windowlet and the SensL G-APD is ≈$6x10^8$, for 25 keV photoelectrons.

A 405 nm LED light source was centered axially 25 cm above the prototype, and pulsed at a frequency of 66.7 kHz. We have used a pair of 5 GHz digital LeCroy oscilloscopes to collect, analyze and plot the data. The data presented in this article were collected exclusively during a stress test run, conducted at the acceleration potential fixed to the highest practical setting (25 kV), and at the highest recommended bias for the G-APD. These extreme conditions are by no means optimal for any practical application.

4. WAVEFORMS AND PULSE HEIGHT SPECTRA

The gain of about $6x10^8$ guarantees unambiguous detection of single photons, possible even with unamplified signals (see Figs. 2-5). Approximately 30-40% of electrons scatter back into vacuum from the LYSO scintillator after a fraction of their energy has already been converted into secondary photons. Some of them shortly return into the Windowlet for their second detection, so their entire energy is eventually recorded. The amplitudes of the non-returning electrons fall below the single-photon peak, see Fig.3. However, most of them are still unambiguously detected thanks to the powerful separation of such events from the pedestal. Moreover, thanks to the low temperature in the IceCube environment, the Dark Count contribution will be strongly suppressed and there will be less confusion even in the narrow overlap region. In cases of multiple photons per event, the combination of fully detected, returning, and non-returning back-scattered electrons leads to events that fall between the individual photon peaks, see Figs. 2-5. The probability for back-scattering is related to the square of the effective atomic number of the scintillator, which for LYSO is very high ($Z_{eff}$ =66). Nevertheless, thanks to the high gain, LYSO provides a very good solution for single-photon applications, see Fig. 5. Some low-Z scintillators can be used instead.

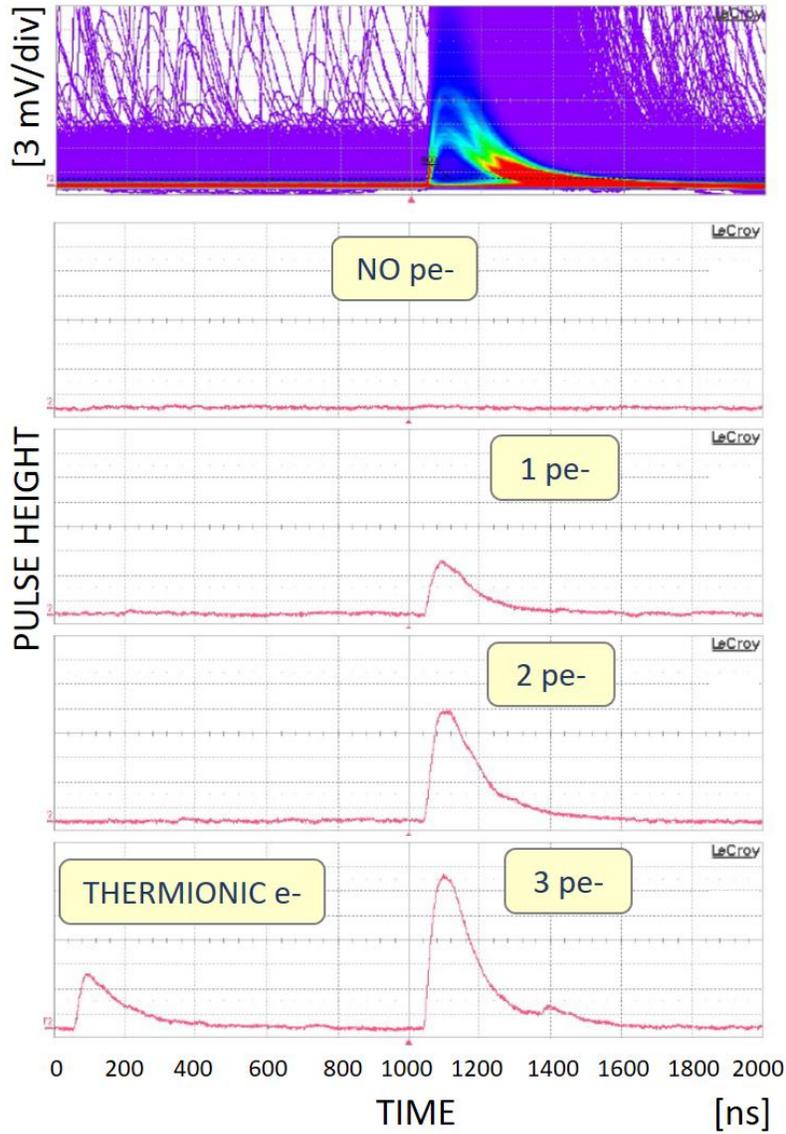

Fig. 2 Screenshots of waveform samples taken directly from the oscilloscope, without any prior amplification, after a 2 m-long cable. Note that the gain is ≈6x10$^8$. The trigger is fixed at 1000 ns in the presented time scale.

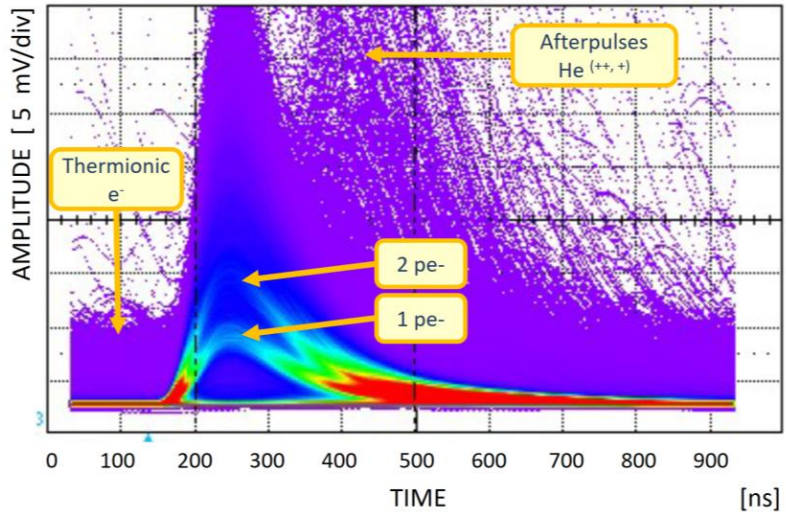

Fig. 3 Superimposed waveforms with different contributions indicated (no amplification was used before the oscilloscope).

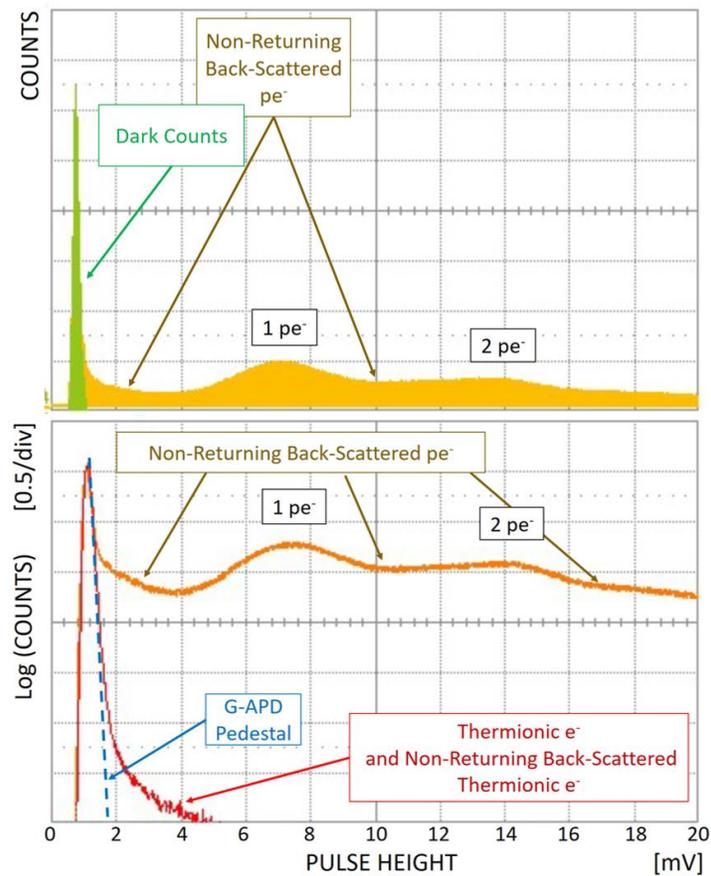

Fig. 4 The spectrum of pulse-height maxima. Top: sampled in the area preceding the peak (green), and in the peak area (yellow). Bottom: the same, but presented in a logarithmic scale (no amplification was used before the oscilloscope).

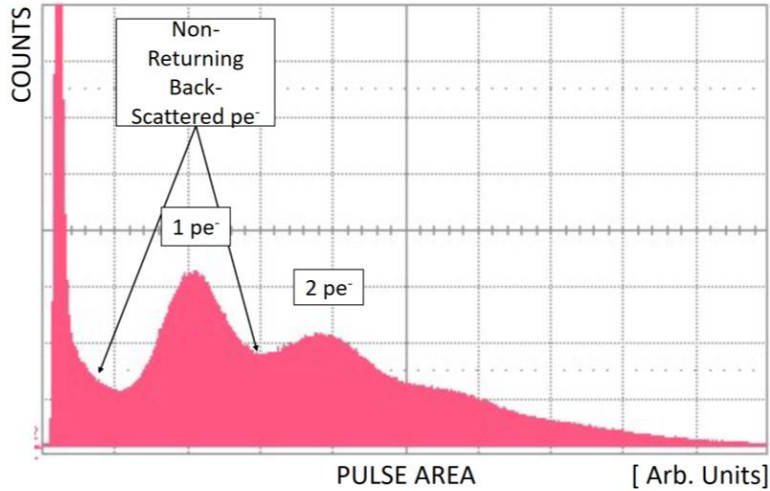

Fig. 5 Distribution of pulse area (no amplification was used before the oscilloscope).

5. AFTERPULSING

The presence of residual gas in any photocathode-based vacuum photosensor is detrimental, because it both leads to the ion-feedback afterpulsing noise, and it harms the photocathode and other surfaces due to ion impacts and chemical contamination. Such afterpulses can be seen in Fig. 3, most of them occurring up to about 200 ns after the main pulse peaks. In a more detailed analysis, we used again the standard output from the SensL G-APD, but this time amplified by a ZX60 preamplifier for better pulse shaping. In addition, the signal was up-scaled from 8 to 10 bit resolution, using oscilloscope's own FIR interpolation routine.

On their way from the photocathode to the Windowlet, a fraction of photoelectrons ionize a neutral atom or a molecule. Each generated ion is then accelerated towards the photocathode, where secondary electrons emerge upon its impact. Their detection leads to the afterpulse, i.e. the "stop" signal used in the ion time of flight (TOF) measurement, while the "start" signal corresponds to the initial photoelectron(s), see Fig. 6. The resulting spectrum of the measured TOF values is shown in Fig. 7. The dominant contribution of the TOF spectrum is from helium, while the contributions of the ions of carbon, nitrogen and oxygen are very weak and fused together.

The spectra of pulse height maxima for both the photoelectron pulses and the afterpulses are shown in Fig. 8. In the selection of events for the TOF analysis, we applied the same threshold on both pulses, equal to one quarter of the one-photon peak value. The afterpulses of a single-electron amplitude form a prominent peak that seems anomalously high compared to the rest of the spectrum. Indeed, apart from the true afterpulses, this spectrum comprises accidental coincidences involving thermionic electrons, which in the TOF spectrum (Fig. 7) correspond to the flat background. Corrected for this effect, the afterpulsing rate is $5 \pm 2 \times 10^{-3}$ ions per photoelectron, which is consistent with the previous measurements of one of the first-generation prototypes [10]. The afterpulsing rate in ABALONE Photosensors is approximately two orders of magnitude lower than in PMTs of a similar size. As elaborated in Ref. [10], that is both because of the exceptional cleanliness of the ABALONE Technology, and—when it comes to the unavoidable and irreducible helium presence—by the two orders of magnitude lower ionization energy-loss rate in the residual gas, thanks to the proportionally higher kinetic energy of electrons in an ABALONE Photosensor [10, 17].

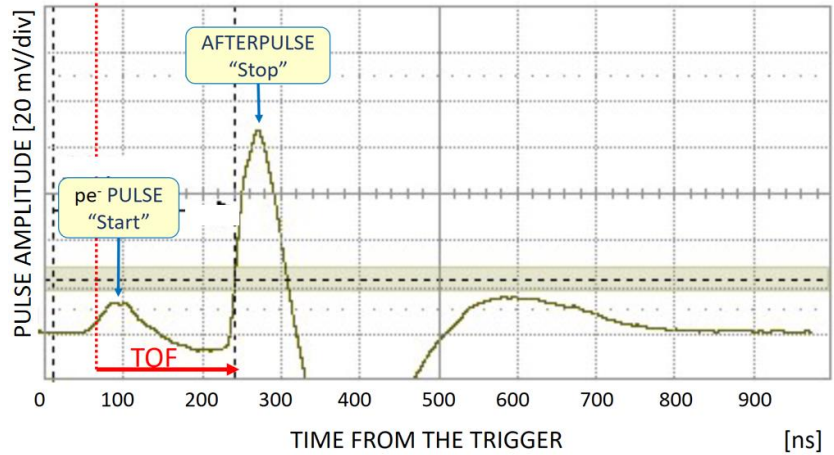

Fig. 6 A typical example of a pulse followed by an afterpulse. The pulse timing has been determined by constant fraction discrimination at 25% of the pulse maximum. The time of flight of the ion that has caused this afterpulse (most likely a $He^+$ ion) is the difference between the times of the two pulses.

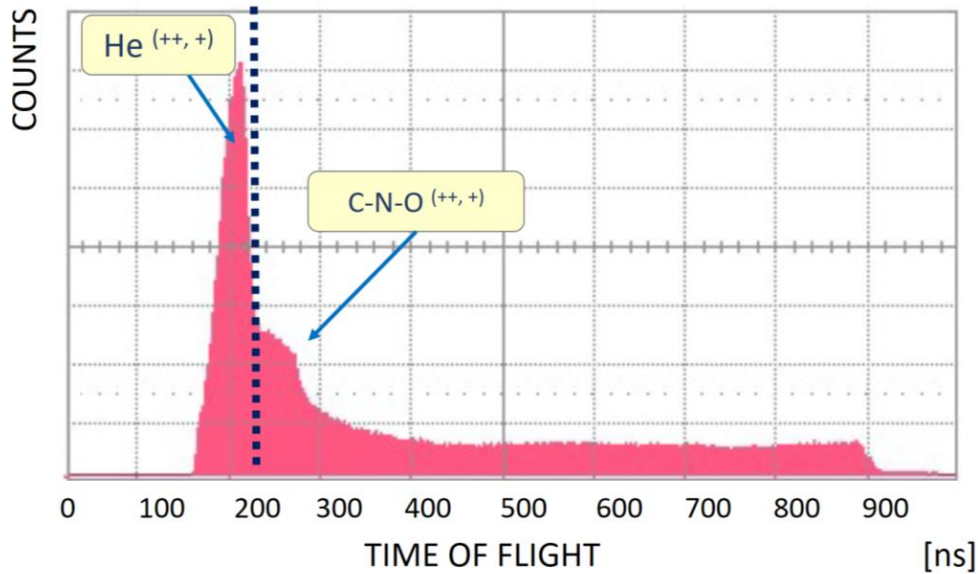

Fig. 7 The spectrum of the time of flight measurements. Helium is the only gas that penetrates glass, and that can be neither chemisorbed, nor permanently implanted. It is therefore the dominant residual gas component in the vacua of ABALONE Photosensors. The vertical dashed line indicates the maximum possible TOF of a $He^+$ ion.

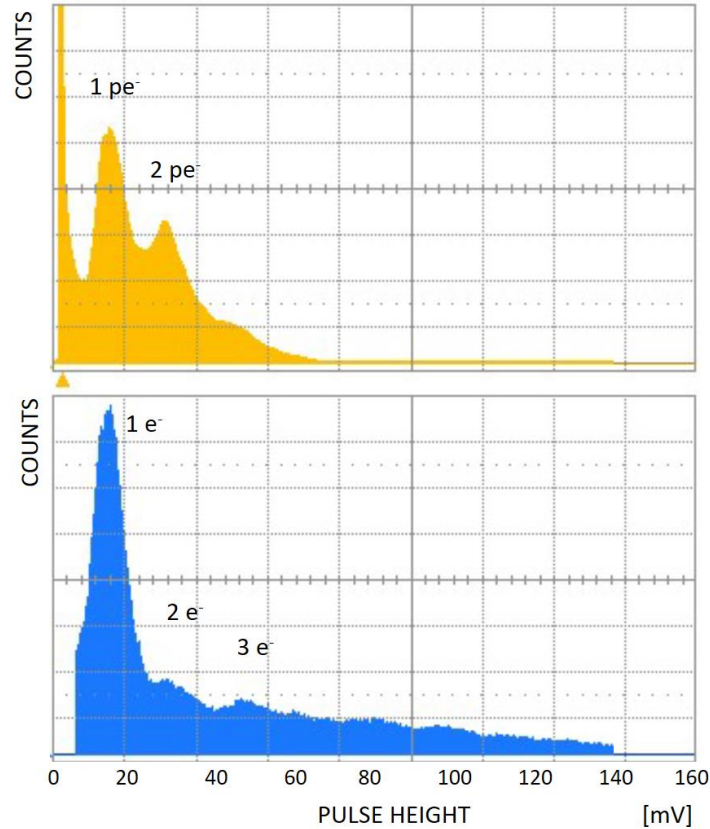

Fig. 8 Spectra of pulse height maxima for the photoelectron pulses (top), and for the afterpulses (bottom).

Each of the following ABALONE design features plays an important role: (a) The low base pressure of $10^{-11}$ Torr in the production system; (b) The absence of outgassing sources, such as metals, ceramics, feedthroughs, microchannel plates, or silicon diodes; (c) The complete absence of electron multiplication, and thus electron-induced desorption within the vacuum; (d) The integrity of the two thin-film alloy seals; (e) Minimum area, open geometry and high molecular conductivity, all necessary for effective pumping. The following three internal vacuum pumping mechanisms keep further improving the vacuum quality in the ABALONE Photosensors over the lifetime: (i) Chemisorption on the multi-functional, chemically reactive thin-film getter that covers about 90% of the Base Plate surface, (ii) Ion implantation in glass behind the photocathode [10, 18,19], and (iii) Physisorption on all passive surfaces, none of which suffers from electron bombardment.

## 6. TIMING RESOLUTION AND DARK CURRENT

The waveforms shown in Fig. 9 were sampled from the fast readout of the SensL G-APD. The signal was attenuated by -3 dB, and then amplified by a fast Ortec VT120C preamplifier (total amplification of 14.1), at the entry to the oscilloscope, after passing through a 2m long 50 Ω SMA cable. Thanks to the high gain of the ABALONE Phosotsensor, all pulses rise sharply, despite the 42 ns-long decay time of the LYSO scintillator. Simple constant-fraction discrimination at 5% of the peak height provides a time distribution in Fig. 10, with a width of 740 ps, fwhm. This width presents an upper limit on the actual timing resolution for a number of reasons, starting with the cable length and the simplistic time discrimination. Furthermore, we have used an ordinary LED pulser, whose own jitter, roughly estimated as 500 ps, is commensurate with the observed effect. Indeed, according to numerical simulations the maximum span of the times of flight of photoelectrons under the studied conditions is between 2.3 ns and 1.5 ns, for photoelectrons originating from the middle and the periphery of the Dome, respectively. The projective geometry of the conical light beam (formed by the point-like LED pulser, positioned 2.5 Dome diameters in front of the Dome) restricts that interval further to about 2.3-1.8 ns, i.e. to an effective fwhm resolution of around 250 ps. The measured time resolution of 740 ps fwhm is thus consistent with all of the contributing effects, and further optimization can provide even a significantly better result.

The dark current measurement leads to 0.2 fA/cm$^2$ (at 25 kV), which is lower than the 1 fA/cm$^2$, expected for the relatively noisy $Cs_3Sb$ photocathode [16]. As expected, lower dark currents were found at lower acceleration potentials, and lower temperatures. The dependence and optimization of the photon detection efficiency, dark current and other performance parameters on the acceleration potential, G-APD bias, area and implementation, as well as the temperature, will be discussed in the forthcoming articles.

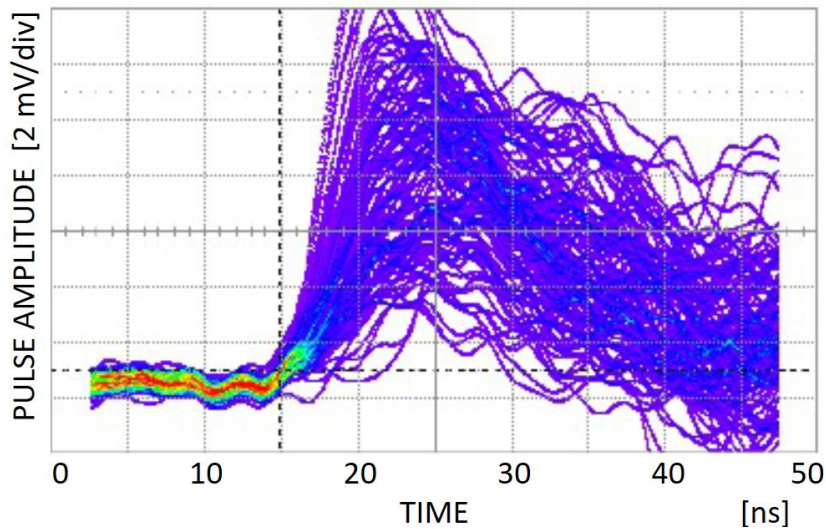

Fig. 9 A sample of superimposed triggered waveforms. The sharp rise of the signals within about one nanosecond is further analyzed in Fig. 10.

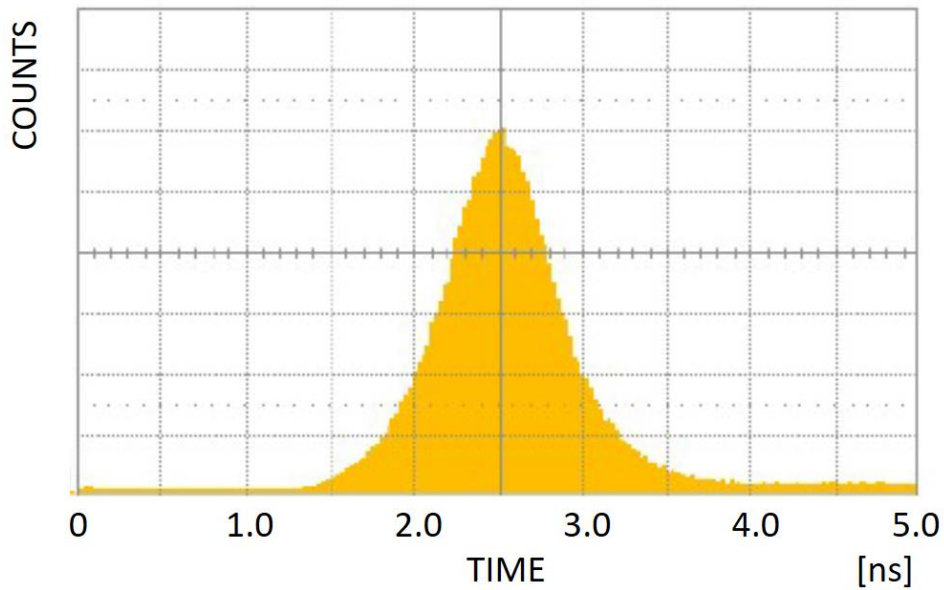

Fig. 10 Distribution of pulse times obtained by constant-fraction timing analysis of fast pulses. The width of 740 ps fwhm includes the genuine timing resolution and instrumental effects such as the LED-pulser jitter.

7. SUMMARY

The presented data were collected during the 120 day-long stress test of the advanced fused-silica ABALONE Phosotesnsor prototypes, specifically designed for the potential large extension of the IceCube cosmic neutrino experiment at the Antarctica. In general, ABALONE Photosensors present a promising solution for all application areas that require cost-effective, robust, and/or modular large-area photosensors, including large experiments in fundamental science, as well as new types of scanners for both functional medical imaging and nuclear security. They are resistant to shock, vibration, compression, low temperatures, and immune to accidental exposure to strong light, including daylight. This modern technology—specifically invented for cost-effective mass production—provides sensitivity to visible and UV light, exceptional radio-purity, and high overall detection performance. The presented performance makes a clear difference: intrinsic gain of $\approx 6\times 10^8$, total afterpulsing rate of only $5\times 10^{-3}$ ions (mostly helium) per photoelectron, sub-nanosecond timing resolution, single-photon sensitivity, as well as unprecedentedly high radio-purity and UV sensitivity, thanks to fused silica components, and at no additional cost to the assembly process.

ACKNOWLEDGMENTS

PhotonLab, Inc. acknowledges Award ID No. 1722351 received from the National Science Foundation. S.B. acknowledges the support of the cluster of excellence PRISMA. A. F. acknowledges support by the Wallenberg Foundation (P.I. J. Conrad).